\documentstyle[pre,epsf,aps,twocolumn]{revtex}	

\newcommand{\C}{{\sf C\hspace*{-0.9ex}\rule{0.15ex}%
{1.3ex}\hspace*{0.9ex}}}

\begin{document} 
\draft
\preprint{EN6375}

\title{Estimating probability densities from short samples: 
	\protect\\
	a parametric maximum likelihood approach} 
\author{T. Dudok de Wit \cite{affil1} and E. Floriani \cite{affil2}}
\date{\today}

\address{Centre de Physique Th\'eorique, Luminy, case 907, \\
	13288 Marseille cedex 9, France}
\date{June 16, 1998}
\maketitle


\begin{abstract}
A parametric method similar to autoregressive spectral estimators
is proposed to determine the probability density function (pdf)
of a random set. The method proceeds by maximizing the
likelihood of the pdf, yielding estimates that perform equally 
well in the tails as in the bulk of the distribution. It
is therefore well suited for the analysis short sets drawn from smooth
pdfs and stands out by the simplicity of its computational
scheme. Its advantages and limitations are discussed.
\end{abstract}


\vspace*{10mm}


\section{Introduction}
\label{sec:introduction}

There are many applications in which it is 
necessary to estimate the probability density
function (pdf) from a finite sample of
$n$ observations $\{x_1, \, x_2, \, \ldots, \, x_n\}$
whose true pdf is $f(x)$. Here we consider the generic
case in which the identically distributed (but not
necessarily independent) random variables have a 
compact support $x_k \in [a,b]$.

The usual starting point for a pdf estimation is the naive
estimate
\begin{equation}
\label{eq:rawpdf}
	\hat{f}_{\delta}(x) = \frac{1}{n}
		\sum_{i=1}^n \delta(x-x_i) \ ,
\end{equation}
where $\delta(.)$ stands for the Dirac delta function.
Although this definition has a number of advantages, it
is useless for practical purposes
since a smooth functional is needed.

Our problem consists in finding an estimate
$\hat{f}(x)$ whose integral over an interval of
given length converges toward that of the true pdf
as $n \rightarrow \infty$. Many solutions have been
developed for that purpose: foremost among these are
kernel techniques in which the estimate $\hat{f}_{\delta}(x)$
is smoothed locally using a kernel function $K(x)$
\cite{Tapia78,Silverman86,Izenman91}
\begin{equation}
\label{eq:kernel}
	\hat{f}(x) = \int_a^b \frac{1}{w} K 
		\left( \frac{x-y}{w} \right) \,
		\hat{f}_{\delta}(y) \, {\rm  d} y \ ,
\end{equation}
whose width is controlled by the parameter $w$.
The well-known histogram is a variant of this technique.
Although kernel approaches are by far 
the most popular ones, the choice of a suitable width
remains a basic problem for which visual guidance
is often needed. More generally, one faces the
problem of choosing a good partition. 
Some solutions include Bayesian 
approaches \cite{Wolpert95}, polynomial fits
\cite{Ventzek94} and methods based on wavelet 
filtering \cite{Donoho93}.

An alternative approach, considered by many authors
\cite{Cenkov62,Schwartz67,Kronmal68,Crain74,Pinheiro97},
is a projection of the pdf on orthogonal functions
\begin{equation}
\label{eq:projection}
	\hat{f}(x) = \sum_k \alpha_k \; g_k(x) \ ,
\end{equation}
where the partition problem is now treated in dual space.
This parametric
approach has a number of interesting properties: 
a finite expansion often suffices to obtain a good 
approximation of the pdf and the convergence 
of the series versus the sample size $n$ is generally
faster than for kernel estimates. A strong point
is its global character, since the pdf is fitted globally, 
yielding estimates that are better behaved in
regions where the lack of statistics causes kernel
estimates to perform poorly. 
Such a property is particularly
relevant for the analysis of turbulent wavefields, in
which the tails of the distribution are of great
interest (e.g. \cite{Frisch95}).

These advantages, however, should be weighed against
a number of downsides. Orthogonal series do not 
provide consistent estimates 
of the pdf since for increasing number of terms they
converge toward $\hat{f}_{\delta}(x)$ instead of
the true density $f(x)$ \cite{Devroye85}.
Furthermore, most series can only handle continuous or 
piecewise continuous densities. 
Finally, the pdf estimates obtained that way
are not guaranteed to be nonnegative (see
for example the problems encountered in \cite{Nanbu95}).

The first problem is not a major obstacle, since
most experimental distributions are smooth anyway.
The second one is more problematic. 
In this paper we show how it can be partly overcome by 
using a Fourier series expansion of the pdf and
seeking a maximization of the likelihood
\begin{equation}
	\hat{L} = \int_a^b 
		\log \hat{f}(x) \; {\rm  d} x \ .
\end{equation}
The problem of choosing an appropriate partition
then reduces to that of fitting the pdf with
a positive definite Pad\'e approximant \cite{Graves72}.

Our motivation for presenting this particular
parametric approach stems from its
robustness, its simplicity and the originality 
of the computational scheme it leads to. The latter,
as will be shown later, 
is closely related to the problem of estimating
power spectral densities with
autoregressive (AR) or maximum entropy methods
\cite{Priestley81,Haykin79,Percival93}.
To the best of our knowledge, the only earlier
reference to similar work is that by
Carmichael \cite{Carmichael76};
here we emphasize the relevance of the method
for estimating pdfs and propose a criterion for choosing
the optimum number of basis functions.


\section{The Maximum likelihood approach}
\label{sec:method}

The method we now describe basically involves
a projection of the pdf on a Fourier series.
The correspondence between the continuous pdf $f(x)$ 
and its discrete characteristic function $\phi_k$
is established by \cite{Oppenheim89}
\begin{eqnarray}
\label{eq:IFT2}
\phi_k &=& \int_{-\pi}^{+\pi} f(x) \; e^{j k x}
		\; {\rm  d} x \\
\label{eq:DFT2}
f(x) &=& 2 \pi \sum_{k=-\infty}^{+\infty}
		\phi_k \; e^{-j k x} \ ,
\end{eqnarray}
where $\phi_k = \phi_{-k}^* \in {\C}$ is hermitian
\cite{Comment1}.
Note that we have applied a linear transformation
to convert the support from $[a,b]$ to $[-\pi, \pi]$.

For a finite sample, an unbiased estimate of the
characteristic function is obtained by inserting
eq.~\ref{eq:rawpdf} into eq.~\ref{eq:IFT2}, giving
\begin{equation}
\label{eq:charfctn}
	\hat{\phi}_k = \frac{1}{n} \sum_{i=1}^{n} e^{jkx_i} \ .
\end{equation}
The main problem now consists in recovering the pdf from
eq.~\ref{eq:DFT2} while avoiding the infinite
summation. By working in dual space we have substituted the
partition choice problem by that of selecting the
number of relevant terms in the Fourier series expansion.

The simplest choice would be to truncate 
the series at a given ``wave number'' $p$ and discard 
the other ones
\begin{equation}
\label{eq:DFT3}
\hat{f}(x) = 2 \pi \sum_{k=-p}^{+p}
		\hat{\phi}_k \; e^{-j k x} \ .
\end{equation}
Such a truncation is equivalent
to keeping the lowest wave numbers and thus filtering
out small details of the pdf.
Incidentally, this solution is equivalent to a kernel
filtering with $K(x) = \sin(\pi x)/\pi x$ as kernel.  
This kernel is usually avoided because it suffers from 
many drawbacks such as the generation of spurious 
oscillations.

An interesting improvement was suggested by Burg in the
context of spectral density estimation
(see for example~\cite{Priestley81,Haykin79}). 
The heuristic idea is to keep some
of the low wave number terms while the remaining ones, 
instead of being set to zero, are left as free parameters:
\begin{eqnarray}
\label{eq:DFT4}
\hat{f}(x) = 2 \pi \sum_{k=-\infty}^{+\infty}
		\hat{\alpha}_k \; e^{-j x k} \\
	\mbox{with} \ \ \hat{\alpha}_k = \hat{\phi}_k,
		 \ \ \ |k| \leq p \nonumber \ .
\end{eqnarray}
The parameters $\hat{\alpha}_k$, for $|k| > p$, are then fixed 
self-consistently according to some criterion.

We make use of this freedom 
to constrain the solution to a particular class of
estimates. Without
any prior information at hand, a reasonable choice is to 
select the estimate that contains the least
possible information or is the most likely. It is therefore
natural to seek a maximization of an entropic quantity
such as the sample entropy
\begin{equation}
	\hat{H} = - \int_{-\pi}^{+\pi} 
		\hat{f}(x) \log \hat{f}(x) \; {\rm  d} x \ ,
\end{equation}
or the sample likelihood
\begin{equation}
\label{eq:likelihood}
	\hat{L} = \int_{-\pi}^{+\pi} 
		\log \hat{f}(x) \; {\rm  d} x \ .
\end{equation}
We are a priori inclined to choose the entropy 
because our objective is the estimation of 
the pdf and not that of the characteristic function. 
However, numerical investigations done in the context of 
spectral density estimation rather lend support to
the likelihood criterion \cite{Johnson84}. 
A different and stronger motivation for preferring
a maximization of the likelihood comes from the
simplicity of the computational scheme it gives rise to.

This maximization means that the tail of the characteristic 
function is chosen subject to the constraint
\begin{equation}
\label{eq:differential}
	\frac{\partial \hat{L}}{\partial \hat{\alpha}_k} = 0,
		 \ \ \ |k|>p \ .
\end{equation}
From eqs.~\ref{eq:DFT4} and \ref{eq:likelihood} the 
likelihood can be rewritten as
\begin{equation}
\label{eq:likelihood2}
	\hat{L} = \int_{-\pi}^{+\pi} \log 
		\left(	2 \pi \sum_{k=-\infty}^{+\infty}
			\hat{\alpha}_k \; e^{-j x k}
		\right)  \, {\rm  d} x \ .
\end{equation}
As shown in the appendix, the likelihood is maximized
when the pdf can be expressed by the functional
\begin{equation}
\label{eq:ck}
	\hat{f}_p(x) = \frac{1}{\sum_{k=-p}^{p} c_k e^{-jkx}} \ ,
\end{equation}
which is a particular case of a  Pad\'e approximant 
with poles only and no zeros  \cite{Graves72}.
Requiring that $\hat{f}_p(x)$ is real and bounded, 
it can be rewritten as
\begin{equation}
\label{eq:pdfestim}	
	\hat{f}_p(x) = \frac{\varepsilon_0}
		{2 \pi} \, \frac{1}{\left|1 + a_1
		e^{-jx} + \cdots + a_p
		e^{-jpx} \right|^2} \ .
\end{equation}
The values of the coefficients $\{a_1,\ldots,a_p\}$ 
and of the normalization constant $\varepsilon_0$ 
are set by the condition that the Fourier transform
of $\hat{f}_p(x)$ must match the sample characteristic
function $\hat{\phi}_k$ for $|k| \le p$.

This solution has a number of remarkable properties,
some of which are deferred to the appendix. Foremost among 
these are its positive definite character and the simple
relationship which links 
the polynomial coefficients $\{a_1,\ldots, a_p\}$ 
to the characteristic function on which they
perform a regression. Indeed, we have
\begin{eqnarray}
\label{eq:ARmodel}
	\hat{\phi}_k + a_1 \hat{\phi}_{k-1} + 
		a_2 \hat{\phi}_{k-2} + \cdots
		+ a_p \hat{\phi}_{k-p} = &0&, 	\\
	1\leq k\leq &p& \nonumber \ .
\end{eqnarray}
This can be cast in a set of  Yule-Walker equations
whose unique solution  contains the polynomial coefficients 
\begin{equation}
\label{eq:Yule}
	\left[ \begin{array}{cccc}
		\hat{\phi}_0	&  \hat{\phi}_{-1}  
			& \cdots 	& \hat{\phi}_{-p+1} \\
		\hat{\phi}_1	&  \hat{\phi}_0     
			& \cdots  	& \hat{\phi}_{-p+2} \\
		\vdots		&  \vdots
			&		& \vdots \\
		\hat{\phi}_{p-1}&\hat{\phi}_{p-2} 
			& \cdots  & \hat{\phi}_0
	\end{array} \right] \;
	\left[ \begin{array}{c}
		a_1 \\ a_2 \\ \vdots \\ a_p
	\end{array} \right] = - 
	\left[ \begin{array}{c}
		\hat{\phi}_1 \\ \hat{\phi}_2 \\ 
		\vdots \\ \hat{\phi}_p
	\end{array} \right] \ .
\end{equation}
Advantage can be taken here of the Toeplitz 
structure of the matrix. The proper normalization 
($\int_{-\pi}^{+\pi} \hat{f}(x) \, {\rm  d} x = 1$) of the pdf is 
ensured by the value of $\varepsilon_0$, which is given by 
a variant of eq.~\ref{eq:ARmodel}
\begin{equation}
	\hat{\phi}_0 + a_1 \hat{\phi}_{-1} + 
		a_2 \hat{\phi}_{-2} + \cdots
		+ a_p \hat{\phi}_{-p} = \varepsilon_0 \ .
\end{equation}
Equations \ref{eq:pdfestim} and \ref{eq:Yule} 
illustrate the simplicity of the method.


\section{Some properties}
\label{sec:properties}

A clear advantage of the method over 
conventional series expansions
is the automatic positive definite character of
the pdf. Another asset is the close
resemblance with autoregressive or maximum entropy methods
that are nowadays widely used in the estimation of spectral 
densities. Both methods have in common
the estimation of a positive function by means of
a Pad\'e approximant whose coefficients directly
issue from a regression (eq.~\ref{eq:ARmodel}). This
analogy allows us to exploit here some results previously
obtained in the framework of spectral analysis.

One of these concerns the statistical properties of the
maximum likelihood estimate. These properties are badly known
because the nonlinearity of the problem impedes 
any analytical treatment. The analogy with spectral densities, 
however, reveals that the
estimates are asymptotically normally distributed
with a standard deviation \cite{Berk74,Parzen74}
\begin{equation}
	\sigma_{\hat{f}} \propto \hat{f} \ .
\end{equation}
This scaling should be compared against that of 
conventional kernel estimates, for which
\begin{equation}
	\sigma_{\hat{f}} \propto \sqrt{\hat{f}} \ .
\end{equation}
The key point is that kernel estimates are relatively
less reliable in low density regions than in the bulk of the
distribution, whereas the relative uncertainty of maximum likelihood
estimates is essentially constant.
The latter property is obviously preferable when the tails 
of the distribution must be investigated, 
e.g. in the study of rare events.

Some comments are now in order. By choosing a Fourier series
expansion, we have implicitly assumed that the pdf was
$2 \pi$-periodic, which is not necessarily the case. Thus special
care is needed to enforce periodicity, since otherwise
wraparound may result \cite{Scargle89}. 
The solution to this problem depends on how easily
the pdf can be extended periodically.
In most applications, the tails of the distribution progressively
decrease to zero, so periodicity may be enforced simply by artificially
padding the tails with a small interval in which the density vanishes.
We do this by rescaling the support from $[a,b]$
to an interval which is slightly smaller than $2 \pi$, say
$[-3, 3]$ \cite{Comment2}. 
Once the Pad\'e approximant is known,
the $[-3, 3]$ interval is scaled back to $[a,b]$.

If there is no natural periodic extension to the pdf,
(for example if $f(a)$ strongly differs from $f(b)$)
then the choice of Fourier basis functions in 
eq.\ \ref{eq:projection} becomes questionable
and, not surprisingly, the quality of the fit
degrades. Even in this case, however, the results
can still be improved by using ad hoc solutions
\cite{Comment3}.

We mentioned before that the maximum likelihood method
stands out by computational simplicity. Indeed, a minimization 
of the entropy would lead to the solution
\begin{equation}
\label{eq:pdfestim2}	
	\log \hat{f}_p(x) \propto 
		\sum_{k=-p}^{p} c_k e^{-jkx} \ ,
\end{equation}
whose numerical implementation requires 
an iterative minimization and is therefore
considerably more demanding.

Finally, the computational cost is found to be
comparable or even better (for large sets) than for 
conventional histogram estimates.
Most of the computation time goes into the
calculation of the characteristic function, for which
the number of operations scales as the sample size $n$.


\section{Choosing the order of the model}
\label{sec:order}

The larger the order $p$ of the model is, the finer the
details in the pdf estimate are. Finite
sample effects, however, also increase with $p$.
It is therefore of
prime importance to find a compromise.
Conventional criteria for selecting the best compromise between
model complexity and quality of the fit,
such as the Final Prediction Error and the Minimum Description
Length \cite{Priestley81,Haykin79,Percival93} are 
 not applicable here because they require the series
of characteristic functions $\{ \phi_k \}$ to be normally
distributed, which they are not.

Guided by the way these empirical criteria have been chosen,
we have defined a new one, which is based on the
following observation: as $p$ increases starting from 0,
the pdfs $\hat{f}_p(x)$ progressively converge toward
a stationary shape; after some optimal order, however, 
ripples appear and the shapes start diverging again.
It is therefore reasonable to compare the
pdfs pairwise and determine how close they are.
A natural measure of closeness between
two positive distributions $\hat{f}_p(x)$ and 
$\hat{f}_{p+1}(x)$ is the Kullback-Leibler entropy
or information gain \cite{Kullback51,Beck93}
\begin{equation}
	\hat{I}(\hat{f}_{p+1},\hat{f}_p) = 
		\int_{-\pi}^{+\pi} \, \hat{f}_{p+1}(x) 
		\, \log \frac{\hat{f}_{p+1}(x)}
		{\hat{f}_p(x)} \; {\rm  d} x \ ,
\end{equation}
which quantifies the amount of information gained
by changing the probability density describing
our sample from $\hat{f}_p$ to $\hat{f}_{p+1}$. 
In other words,
if $H_p$ (or $H_{p+1}$) is the hypothesis that $x$ 
was selected from the population whose probability
density is $\hat{f}_p$ ($\hat{f}_{p+1}$), then
$\hat{I}(\hat{f}_{p+1},\hat{f}_p)$
is given as the mean information for discriminating 
between $H_{p+1}$ and $H_{p}$ per observation 
from $\hat{f}_{p+1}$ \ \cite{Kullback51}.

Notice that the information gain is not a distance
between distributions; it nevertheless has
the property of being non negative and to
vanish if and only if
$\hat{f}_p \equiv \hat{f}_{p+1}$. We now proceed as 
follows : starting from $p=0$ the order is 
incremented until the information gain
reaches a clear minimum; this corresponds, as it has
been checked numerically, to the convergence toward
a stationary shape; the corresponding order
is then taken as the requested compromise. Clearly,
there is some arbitrariness in the definition of a
such a minimum since visual inspection and common 
sense are needed. In most cases, however, the solution
is evident and the search can be automated.
Optimal orders usually range between
2 and 10; larger values may be needed to model 
discontinuous or complex shaped densities.


\section{Some examples}
\label{sec:examples}

Three examples are now given in order to illustrate 
the limits and the advantages of the method.


\subsection{General properties}

\begin{figure}
\epsfbox{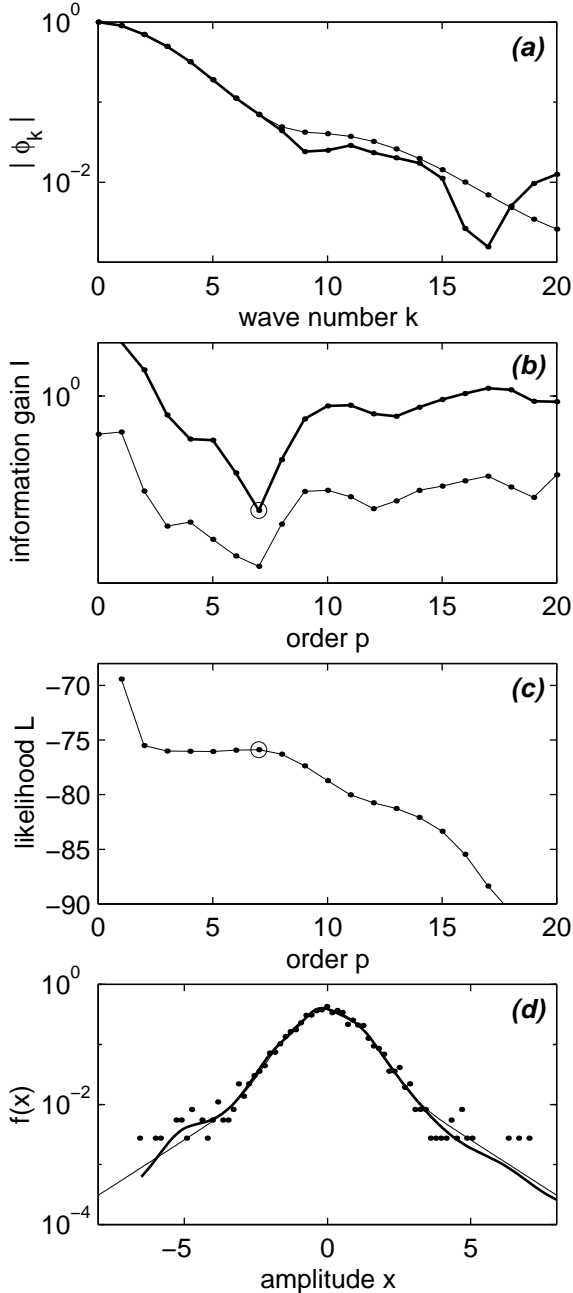}
\caption{Example of a normal distribution with 
exponential tails. The sample size is $n=2000$. From
top to bottom are shown:
(a) the magnitude $|\hat{\phi}_k|$ of the characteristic
function (thick line) and the fit resulting from an 7'th
order model; (b) the information gain (thick line)
and the MISE, both showing a minimum 
around $p=7$ which is marked by a circle; 
(c) the likelihood $\hat{L}$
associated with the different pdfs estimated for $p=$1--20;
and finally (d) the maximum likelihood estimate of the 
pdf (thick line), an estimate based on a histogram 
with 101 equispaced bins (dots) and the true pdf 
(thin line).}
\label{figpdf1}
\end{figure}

First, we consider a normal
distribution with exponential tails
as often encountered in turbulent
wavefields.
We simulated a random sample with $n=2000$ elements 
and the main results appear in Fig.~\ref{figpdf1}. 

The information gain 
(Fig.~\ref{figpdf1}b) decreases as
expected until it reaches a well defined
minimum at $p=7$, which therefore sets
the optimal order of our model. Since the true
pdf is known, we can test this result
against a common measure of the quality of
the fit, which is the Mean Integrated Squared Error
(MISE)
\begin{equation}
	{\rm  MISE}(p) = \int_a^b [ f(x)-\hat{f}_p(x) ]^2  
		{\rm  d} x \ .
\end{equation}
The MISE, which is displayed in
Fig.~\ref{figpdf1}b, also reaches a minimum at
$p=8$ and thus supports the choice of the 
information gain as a reliable indicator for
the best model. Tests carried out on other types of 
distributions confirm this good agreement.

Now that the optimum pdf has been found, its 
characteristic function can be computed and 
compared with the measured one, 
see Fig.~\ref{figpdf1}a. As expected,
the two characteristic functions coincide
for the $p$ lowest wave numbers (eq.~\ref{eq:ARmodel}); 
they diverge at higher wave numbers, for
which the model tries to extrapolate the characteristic
function self-consistently. The fast falloff of the
maximum likelihood estimate explains the
relatively smooth shape of the resulting pdf.

Finally, the quality of the pdf can be
visualized in Fig.~\ref{figpdf1}d,
which compares the measured pdf
with the true one, and an estimate
based on a histogram with 101 bins. An excellent
agreement is obtained, both in the bulk of the
distribution and in the tails, where the
exponential falloff is correctly reproduced. 
This example illustrates the ability
of the method to get reliable estimates in regions 
where standard histogram approaches have a lower
performance.


\subsection{Interpreting the characteristic function}
\label{subsec:interpretation}

\begin{figure}
\epsfbox{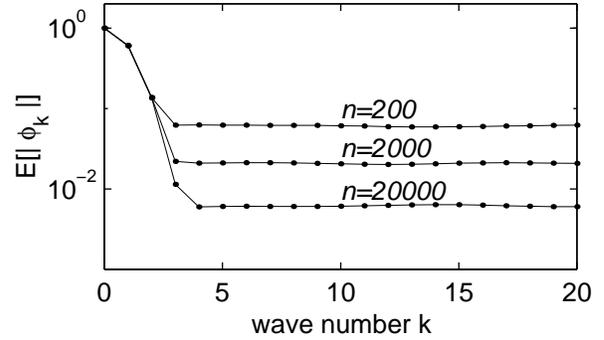}
\caption{The expectation ${\rm  E}[|\hat{\phi}_k|]$
computed for sets of various sizes taken from 
the same normal distribution. The noise-induced
bias level goes down as the size increases, progressively
revealing finer details of the pdf.}
\label{figpdf2}
\end{figure}

The shape of the characteristic function in 
Fig.~\ref{figpdf1}a is reminiscent of spectral
densities consisting of a low wave number (band-limited)
component embedded in broadband noise. A straightforward 
calculation of the expectation of $|\phi_k|$ indeed
reveals the presence of a bias which is due to 
the finite sample size
\begin{equation}
	{\rm  E} [ |\hat{\phi}_k| ] = |\phi_k| + 
	\frac{\gamma }{\sqrt{n}} \ ,
\end{equation}
where $\gamma$ depends on the degree of independence
between the samples in $\{x\}$.
This bias is
illustrated in Fig.~\ref{figpdf2} for independent
variables drawn from a normal distribution, 
showing how the wave number resolution gradually
degrades as the sample size decreases. Incidentally,
a knowledge of the bias level could be used
to obtain confidence intervals for the pdf estimate. This
would be interesting insofar no assumptions have
to be made on possible correlations in the data set.
We found this approach, however, to be too inaccurate 
on average to be useful.

The presence of a bias also gives an
indication of the smallest
scales (in terms of amplitude of $x$) one can reliably 
distinguish in the pdf. For a set of 2000 samples
drawn from a normal distribution, for example,
components with wave numbers in excess of
$k=3$ are hidden by noise and hence the
smallest meaningful scales in the pdf are 
of the order of $\delta x=0.33$. 
These results could possibly be further
improved by Wiener filtering.


\subsection{Influence of the sample size}

To investigate the effect of the sample length $n$, we
now consider a bimodal distribution consisting of two
normal distributions with different means and standard 
deviations. Such distributions are known to be difficult 
to handle with kernel estimators.

\begin{figure}
\epsfbox{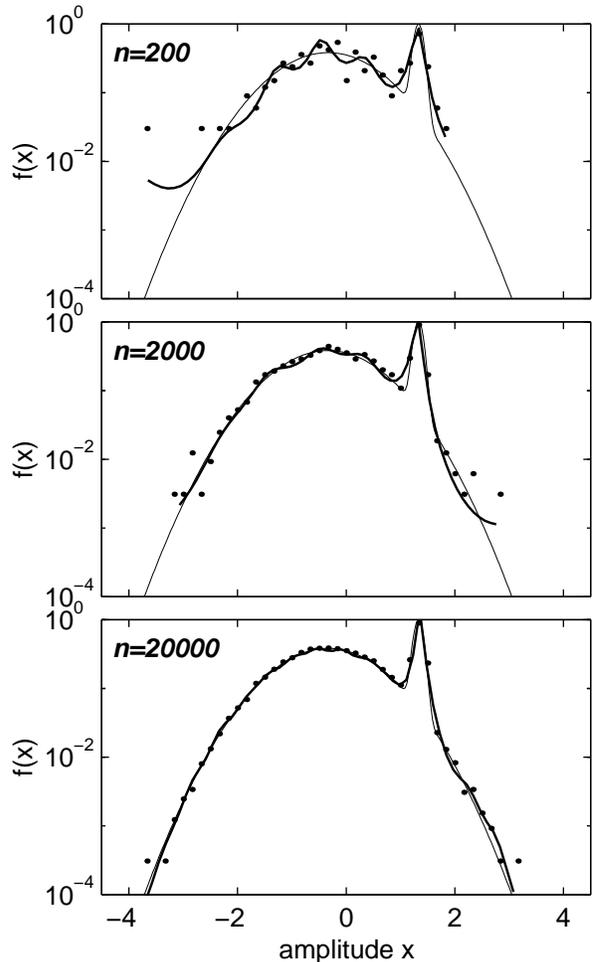}
\caption{The pdfs as calculated for
sets of various sizes taken from the same bi-normal
distribution. The thick line designates the maximum likelihood
estimate, the thin line the true pdf and the dots a
histogram estimate obtained from 61 equispaced bins.
The optimum orders are respectively from top to bottom $p=5$,
$p=6$ and $p=11$.}
\label{figpdf3}
\end{figure}

Samples with respectively $n=200$, $n=2000$
and $n=20000$ elements were generated; their
characteristic functions and the resulting pdfs are
displayed in Fig.~\ref{figpdf3}. Clearly, finite
sample effects cannot be avoided for small samples
but the method nevertheless succeeds relatively
well in capturing the true pdf and in particular the
small peak associated with the narrow distribution. An
analysis of the MISE shows that it is
systematically lower for maximum likelihood
estimates than for standard histogram estimates,
supporting the former.


\subsection{A counterexample}

The previous examples gave relatively good results
because the true distributions were rather smooth.
Although such smooth distributions are
generic in most applications it may be instructive to look at
a counterexample, in which the method fails.

\begin{figure}
\epsfbox{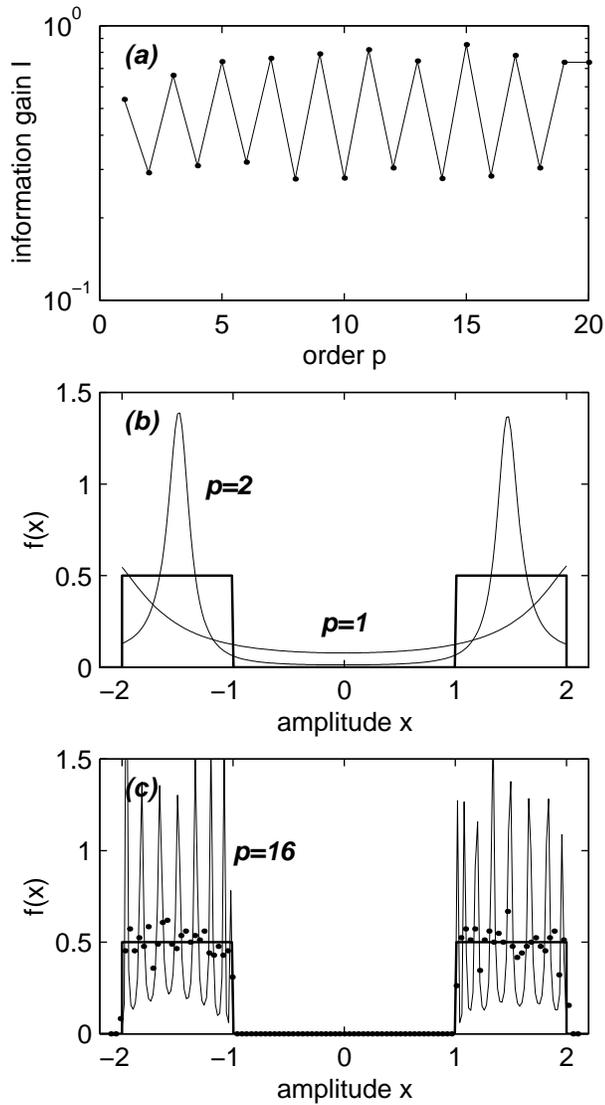}
\caption{Results obtained for an annular distribution;
the sample size is $n=2000$.
In (a) The information gain has no clear minimum and
hence there is no well defined order for the model.
In (b) the estimated pdfs for $p=1$ and $p=2$ fail to 
fit the true pdf (thick line). Increasing the order (c)
improves the fit but also enhances the Gibbs phenomenon.
Dots correspond to a histogram estimate with equispaced bins.}
\label{figpdf4}
\end{figure}

Consider the distribution which
corresponds to a cut through an annulus
\begin{equation}
	f(x) = \left\{ \begin{tabular}{cl}
		$\frac{1}{2}$	& $1 \leq |x| \leq 2$	\\
		0		& elsewhere 		\\
	\end{tabular} \right. \ .
\end{equation}
A sample was generated with $n=2000$ elements
and the resulting information gains are shown
in Fig.~\ref{figpdf4}. There is an ambiguity 
in the choice of the model order and indeed the
convergence of the pdf estimates toward the true
pdf is neither uniform nor in the mean.
Increasing the order improves the fit of the 
discontinuity a little but also increases the 
oscillatory behavior known as the 
Gibbs phenomenon. This problem is related to the
fact that the pdf is discontinuous and hence
the characteristic function is not 
absolutely summable.

Similar problems are routinely encountered in the 
design of digital filters,
where steep responses cannot 
be approximated with infinite impulse response filters
that have a limited number of poles 
\cite{Oppenheim89}. The bad performance of the 
maximum likelihood approach in this case also comes 
from its inability to handle densities that vanish
over finite intervals. A minimization of the
entropy would be more appropriate here.


\section{Conclusion}
\label{sec:conclusion}

We have presented a parametric procedure for
estimating univariate densities using a positive definite
functional. The method proceeds by maximizing the
likelihood of the pdf subject to the constraint that the
characteristic functions of the sample and estimated pdfs 
coincide for a given number of terms. Such
a global approach to the estimation of pdfs is in 
contrast to the better known local methods (such as
non-parametric kernel methods) whose performance is poorer in
regions where there is a lack of statistics, such as the tails
of the distribution. This difference makes the maximum 
likelihood method relevant for the analysis of short
records (with typically hundreds or thousands of samples).
Other advantages include a simple computational
procedure that can be tuned with
a single parameter. An entropy-based criterion has been
developed for selecting the latter.

The method works best with densities that are at least once
continuously differentiable and that can be extended periodically.
Indeed, the shortcomings of the method are essentially the same
as for autoregressive spectral estimates, which
give rise to the Gibbs phenomenon if the density is discontinuous.

The method can be extended to multivariate densities,
but the computational procedures are not yet within the
realm of practical usage.
Its numerous analogies with the design of digital filters
suggest that it is still open to improvements.


\acknowledgments

We gratefully acknowledge the dynamical systems team at the
Centre de Physique Th\'eorique for many 
discussions as well as D. Lagoutte and B. Torr\'esani for
making comments on the manuscript. E. Floriani acknowledges
support by the EC under grant \mbox{nr. ERBFMBICT960891}.


\appendix
\section*{}
\label{sec:appendix}

We detail here the main stages that lead to the pdf estimate
described in Sec.~\ref{sec:method} because extensive proofs
are rather difficult to find in the literature.

The maximum likelihood condition (eq.~\ref{eq:differential}) can
be expressed as
\begin{equation}
	\int_{-\pi}^{+\pi} \frac{e^{-jkx}}{\sum_{l=-\infty}^{\infty} 
		\hat{\alpha}_l e^{-jlx}} \, {\rm  d} x = 
        \int_{-\pi}^{+\pi} \frac{e^{-jkx}}{\hat{f}(x)} \, 
		{\rm  d} x = 0 \ ,
\end{equation}
for $|k| > p$  \cite{Comment4}. 
This simply means that the Fourier expansion of 
$\left(\hat{f}(x)\right)^{-1}$ should not
contain terms of order $|k| > p$ and hence the solution 
must be
\begin{equation}
\label{eq:solution}
	\hat{f}_p(x) = \frac{1}{\sum_{k=-p}^{p} c_k e^{-jkx}} \ .
\end{equation}
The pdf we are looking for must of course be real, 
and so the coefficients should be hermitian
$c_k = c_{-k}^{*}$. We also want the pdf to be 
bounded, which implies that 
\begin{equation}
\label{eq:nocircle}
	\sum_{k=-p}^{p} c_k e^{-jkx} \neq 0 \  , 
		\ \ \ \forall x \in [-\pi,\pi] \ .
\end{equation}
Let us now define, for $z$ complex
\begin{equation}
	C(z) = \sum_{k=-p}^{p} c_{-k} z^k \ ,
\end{equation}
and
\begin{equation}
	P(z) = z^p C(z) \ .
\end{equation}
$P(z)$ is a polynomial of degree $2p$. It can be 
easily verified that \cite{Smylie73}
\begin{equation}
	P(z) = z^{2p} \left[P\left(\frac{1}{z^*}\right)\right]^* \,
\end{equation}
as a consequence of the hermiticity of the coefficients $c_k$.
In particular, this tells us that if $z_1$ 
is a root of $P(z)$, then 
$1/z_1^*$ (the complex-conjugate of its mirror image with 
respect to the unit circle) is also a root of $P(z)$.
From eq.~\ref{eq:nocircle} we know that none of these
roots are located on the unit circle.

Let us now rearrange the roots of $P(z)$,
denoting by $\{z_1,\ldots,z_p\}$ the $p$ roots
lying outside the unit disk and by
$\{1/z_1^*,\ldots,1/z_p^* \}$ the $p$ other
ones that are located inside the unit circle. 
We can then write:
\begin{equation}
	P(z) = c_{-p}(z-z_1)\cdots(z-z_p)
		\left(z-\frac{1}{z_1^*}\right)
		\cdots\left(z-\frac{1}{z_p^*}\right) \ ,
\end{equation}
with
\begin{equation}
	c_{-p} z_1 \cdots z_p = c_{p} z_1^* \cdots z_p^* \ .
\end{equation}
From this $C(z)$ can be written as:
\begin{equation}
	C(z) = \pm B(z)\left[B\left(\frac{1}{z^*}
		\right)\right]^* \ ,
\end{equation}
where
\begin{equation}
	B(z)= \left| \frac{c_p}{z_1\cdots z_p} 
		\right|^{\frac{1}{2}}(z-z_1)\cdots(z-z_p) \ .
\end{equation}
By construction, all the roots of  
$B(z)$ are located outside the unit disk.

Finally, we get for $\hat{f}_p(x)$:
\begin{equation}
\label{eq:B}
	\hat{f}_p(x) = \frac{1}{C(z=e^{jx})} = 
	\pm \frac{1}{\left|B(e^{jx})\right|^2} \ .
\end{equation}
All the solutions of the maximum likelihood principle,
if real and bounded, are thus of constant sign and 
have the structure given by eq.~\ref{eq:B}. Excluding
negative definite solutions we obtain
\begin{equation}
	\hat{f}_p(x) = \frac{\varepsilon_0}
		{2 \pi} \, \frac{1}{\left|1 + a_1
		e^{-jx} + \cdots + a_p
		e^{-jpx} \right|^2} \ ,
\end{equation}
where
\begin{equation}
\varepsilon_0 = \frac{2\pi}{|b_0|^2} \ , \ \ \ 
	a_i = \frac{b_i^*}{b_0^*} \ , \ \ \ i = 1,\cdots,p \ ,
\end{equation}
where $\{b_1,\ldots,b_p\}$ are the coefficients of the
polynomial $B(z)$ and $A(z) = 1 + a_1 z + \cdots + a_p z^p$  
has all its roots outside the unit disk. The normalization
constant $\varepsilon_0$ is set by the condition
\begin{equation}
	\int_{-\pi}^{+\pi} \hat{f}_p(x) \, {\rm  d} x = 1 \ .
\end{equation}
The coefficients $\{a_1,\ldots,a_p\}$ are now identified 
on the basis that the characteristic function of the pdf
estimate $\hat{f}_p(x)$ should match the first $p$ 
terms of the sample characteristic function exactly, namely
\begin{equation}
	\hat{\alpha}_k = \hat{\phi}_k = 
		\int_{-\pi}^{+\pi} \hat{f}_p(x) \; e^{j k x} 
		{\rm  d} x \ , \ \ \ 1 \leq k \leq p \ .
\end{equation}
To this purpose, let us compute the quantity 
$\sum_{k=0}^p a_k \hat{\alpha}_{l-k}$.
Recalling that $A(z)$ is analytic in the unit circle and 
making use of Cauchy's residue theorem, we obtain
\begin{eqnarray}
	\label{eq:1p}
	&\sum_{k=0}^p& a_k \hat{\phi}_{l-k} = 0 \ ,
		 \ \ 1 \leq l \leq p  \\
	\label{eq:eps0}
	&\sum_{k=0}^p& a_k \hat{\phi}_{-k} = \varepsilon_0 \ .
\end{eqnarray}
Equation~\ref{eq:1p} fixes the values of $\{a_1,\ldots,a_p\}$ 
and gives the Yule-Walker equations (eq.~\ref{eq:Yule}).
The solution is unique provided that
\begin{equation}
	\det\left[ \begin{array}{ccc}
		\hat{\phi}_0	 & \cdots  & \hat{\phi}_{-p+1} \\
		\vdots		 &	   & \vdots \\
		\hat{\phi}_{p-1} & \cdots  & \hat{\phi}_0
	\end{array} \right] \neq 0 \ .
\end{equation}
The latter condition is verified except when a repetitive
pattern occurs in the characteristic function. In this happens
then the order $p$ should simply be chosen to be less than the
periodicity of this pattern. 

Besides its positivity, the solution we obtain has a number
of useful properties. First, note that all the terms of
its characteristic function can be computed recursively by
\begin{equation}
\label{eq:statespace}
	\left[	\begin{array}{c}
		\hat{\alpha}_{k+1} \\
		\hat{\alpha}_k \\
		\vdots \\
		\hat{\alpha}_{k-p+2}
		\end{array} \right] \, = \,
	\left[	\begin{array}{cccc}
		-a_1 & -a_2 & \cdots  & -a_p \\
		1 & 0 & \cdots  & 0 \\
		0 & 1 & \cdots  & 0\\
		\vdots & \vdots & & \vdots\\
		0 & 0 & \cdots & 0
		\end{array} \right] \,
	\left[	\begin{array}{c}
		\hat{\alpha}_k \\
		\hat{\alpha}_{k-1} \\
		\vdots \\
		\hat{\alpha}_{k-p+1}
		\end{array} \right] \, ,
\end{equation}
in which the starting condition is set by the $p$ first
values of $\hat{\phi}_k$. From this recurrence
relation the asymptotic behavior of $\hat{\phi}_k$ as
$k \rightarrow \infty$ can be probed by diagonalizing
the state space matrix in eq.~\ref{eq:statespace}. The 
eigenvalues of this matrix are the 
roots $\{1/z_1^*,\cdots,1/z_p^* \}$ (called poles), 
which by construction are all inside the unit disk. 
Therefore
\begin{equation}
	\lim_{k \rightarrow \infty} |\phi_k| \sim 
		e^{\lambda k} \ ,
\end{equation}
where $\lambda$ is related to the largest root and
is always negative since
\begin{equation}
	\lambda = \max_k \, \log \left| \frac{1}{z_k^*} \right|
		 \ \ < \ \ 0 \ .
\end{equation}
This exponential falloff of the characteristic function
explains why the resulting pdf is relatively smooth.

Now that we have found a solution in terms of a $[0,p]$ Pad\'e 
approximant, it is legitimate to ask
whether a $[q,p]$ approximant of the type
\begin{equation}
	\hat{f}_{q,p}(x) = \frac{\left|d_0 + d_1 e^{-jx} 
		+ \cdots + d_q e^{-jqx} \right|^2}
		{\left|1 + a_1 e^{-jx} 
		+ \cdots + a_p e^{-jpx} \right|^2}
\end{equation}
could not bring additional flexibility and hence
provide a better estimate of the pdf. Again, we
exploit the analogy with spectral density estimation,
in which the equivalent of $[q,p]$ Pad\'e approximants are
obtained with autoregressive moving average (ARMA) models.
The superiority of ARMA over AR models is generally agreed
upon \cite{Comment5}, although the MISE does not 
firmly establish it \cite{Haykin79}. Meanwhile we note that
there does not seem to exist a simple variational principle,
similar to that of the likelihood maximization, which 
naturally leads to a $[q,p]$ Pad\'e approximant of the pdf.



\newpage

\begin{references}
\bibitem{Comment1}
This is strictly speaking not the characteristic
function, but its complex conjugate.
\bibitem{Comment2}
The width of the interval actually has little impact on the
outcome as long as it is much less than $2 \pi$. The width
and its asymmetry could possibly be tailored to the pdf
in order to improve the fit a little.
\bibitem{Comment3}
The simplest one consists in symmetrizing the pdf:
the initial sample $\{x_1, \, x_2, \, \ldots, \, x_n\}$ is 
transformed into a new one which is twice a large 
$\{x_1, \, \ldots, \, x_n, 2a-x_1, \, \ldots, \, 2a-x_n\}$, and
covers the interval $[2a-b, b]$. The pdf of this
new sample is computed and one
half of it is kept to obtain the desired result.
This procedure doubles the volume of data, but
the computational cost remains approximately the same
since one half only of the data is actually needed to
estimate the characteristic function.
\bibitem{Comment4}
It can be readily verified that this solution
indeed corresponds to a maximum.
\bibitem{Comment5}
The main advantage of ARMA models is their ability to
model spectral densities that vanish locally.
\end{references}
\end{document}